\documentclass{article}

\usepackage{amssymb,amsfonts,amsmath,stmaryrd}
\usepackage{cite,enumerate,float,indentfirst}
\usepackage{color}
\usepackage{tikz}
\usetikzlibrary{arrows,snakes,backgrounds}

\def\be{\begin{eqnarray}}
\def\ee{\end{eqnarray}}
\def\nn{\nonumber}

\def\p{\partial}
\def\tr{{\rm tr}\,}

\definecolor{red}{rgb}{1,0,0}
\definecolor{orange}{rgb}{1,0.5,0}
\definecolor{violet}{rgb}{0.7,0,1}

\hoffset= - 1.0in         

\begin{document}

\begin{center}
\begin{small}
\hfill ITEP/TH-22/21\\
\hfill IITP/TH-17/21\\
\hfill MIPT/TH-14/21\\
\hfill \\
\end{small}
\end{center}

\begin{center}
\begin{small}
\hfill {\bf to Andrei Mironov \\
\hfill on his 60's birthday}\\
\end{small}
\end{center}

\vspace{.5cm}

\begin{center}
\begin{Large}\fontfamily{cmss}
\fontsize{17pt}{27pt}
\selectfont
	\textbf{A new kind of anomaly: on W-constraints for GKM}
	\end{Large}
	
\bigskip \bigskip

\begin{large}A.Morozov
 \end{large}
\\
\bigskip

\begin{small}
 {\it ITEP Moscow 117218, Russia}\\
  {\it IITP, Moscow 127994, Russia}\\
  {\it MIPT, Dolgoprudny, 141701, Russia}
\end{small}
 \end{center}

\bigskip

\begin{abstract}

We look for the origins of the single equation,
which is a peculiar combination of W-constrains,
which provides the non-abelian W-representation for generalized Kontsevich model (GKM),
i.e. is enough to fix the partition function unambiguously.
Namely we compare it with the scalar projection of the matrix Ward identity.
It turns out that, though similar, the two equations do not coincide,
moreover, the latter one is non-polynomial in time-variables.
This discrepancy disappears for the cubic model if partition function is reduced
to depend on odd times (belong to KdV sub-hierarchy of KP),
but in general such reduction is not enough.
We consider the failure of such direct interpretation of the "single equation"
as a new kind of anomaly, which should be explained and eliminated
in the future analysis of GKM.
\end{abstract}

\bigskip

\section{Introduction }

This is going to be a rather technical paper, targeted at clarification
of the long-standing puzzles of Generalized Kontsevich model (GKM) \cite{Kon}--\!\cite{Zhou}.
It is not fully successful, still it can attract attention to potentially important
aspects of the story.
No doubt, at technical level these observations are well known to people who worked with GKM,
but we make an attempt to summarize them and promote to a more conceptual level.
This is needed because of the new accents introduced into the GKM theory quite recently,
in \cite{MMhl} and \cite{MMMR1}--\!\cite{MMMish}, and the need to explain the origins and the form
of the "single equation" \cite{MMMR2} and the character expansion in terms of Hall-Littlewood polynomials
\cite{MMhl}.
We do not achieve these goals in the present paper, but we try to better explain
the difficulties of one particularly promising suggestion from \cite{MMMish} -- 
with the hope that it gains attention and will be somehow resolved in the near future.

GKM \cite{GKM} is an eigenvalue matrix model \cite{UFN3} with the partition function 
\be
{\cal Z}_V[L] = \int_{N\times N} dX e^{-\tr (V(X) + LX)} =
\frac{e^{\tr M V'(M) - V(M)}}{\sqrt{\det V''(M)}}\cdot Z_V\{p_k\}
:= {\cal Z}^{cl}_V[M] \cdot Z_V\{p_k\}
\label{GKMdef}
\ee
with $V'(M)=L$, which depends on the matrix variable $L$
and satisfies the obvious matrix-valued Ward identity \cite{MMM,GN}
\be
\left\{V'\left(\frac{\p}{\p L^{tr}}\right) - L\right\}{\cal Z}_V[L] = 0
\label{GN}
\ee
We further restrict attention to the monomial case, $V(X) = \frac{X^{r+1}}{r+1}$
and label ${\cal Z}_r$  and $Z_r$ by integer $r$.
Qualitatively the properties of monomial GKM are well known \cite{GKM,UFN3}:

1) the "quantum" pieces $Z_r$ are KP $\tau$-functions of the "time variables"
$p_k = \tr M^{-k}$ (which are $r$-dependent in terms of $L$, $p_k = \tr L^{-k/r}$),

2) they are independent of all $p_{kr}$ and belong to the $r$-reduction of KP \cite{FKN},

3) the {\it shape} of $Z_r\{p\}$ is independent of the size $N$ of the matrix $M$,
only the {\it locus} $p_k = \tr M^{-k}$ where the particular integral is actually defined,
depends on $N$,

4) the "classical" pieces ${\cal Z}^{cl}_r[M]$ also can be expressed through $p_k$,
but {\it their} shapes do depend on $N$ -- this was the reason why these formulas 
are not very popular, and we discuss them in a special section \ref{class} below,

5) Ward identities (\ref{GN}) can be rewritten as an infinite set of $W$-constraints on $Z_r\{p_k\}$
\cite{FKN,GKM,UFN3},
\be
\hat W^{(i+1)}_{nr-r+i} \cdot Z_r = 0, \ \ \ \ i=1,\ldots,r-1, \ \ \ n\geq 1
\label{wcons}
\ee
and, as established recently,

6) $Z_r$ has a peculiar non-Abelian $W$-representation \cite{Wrep,MMMish},
i.e. can be unambiguously described by a {\it single} combination of $W$-constraints,
\be
{\bf Single \ Equation \ (SE):} \ \ \ \ \ \ \ \ \ \
\boxed{
\sum_{i=1}^{r-1} (-)^i \sum_n p_{nr-r+i} \hat W^{(i+1)}_{n-i-1} \cdot Z_r\{p\} = 0
}
\ \ \ \ \ \ \ \ \ \ \ \ \ \ \ \ \ \ \ \ \ \ \ \ \ \ \ \ \ \
\label{singleq}
\ee
nicknamed "single equation" (SE) in what follows,

7)  $Z_r$ possesses character expansion in terms of Hall-Littlewood polynomials \cite{MMhl}.

\bigskip

While rather well established in the case of the ordinary cubic ($r=2$) Kontsevich model \cite{Kon},
these issues are quite difficult to address for $r>2$.
It is the purpose of this paper to make one more technical step in this direction.

Namely,
we study a {\it scalar} implication of {\it matrix} Gross-Newmann equation (\ref{GN}),
\be
{\bf Main \ Equation \ (ME):} \ \ \ \ \ \ \ \
\boxed{
\tr \left( M\left\{\left(\frac{\p}{\p L^{tr}}\right)^r - L\right\}\right){\cal Z}_r[L] = 0
}
\ \ \ \ \ \ \ \ \ \ \ \ \ \ \ \ \ \ \ \ \ \ \ \ \ \ \ \ \ \
\label{maineq}
\ee
and the suggestion of \cite{MMMish} to use it as SE -- a basic equation
which provides $Z_r$  as a unique solution in the form of the non-Abelian $W$-representation.
In this paper we call it "the main equation", or just ME to simplify the reference.
In other words, the main question of the present paper is if
\be
{\bf ME} \ \stackrel{?}{=}\ {\bf SE}
\ee
and, if not, what is the difference.
 
Substitution of a badly controlled system of matrix equations (\ref{GN}),
which is {\it believed}
to be equivalent to an infinite set of $W$-constraints (\ref{wcons}),
by a single equation SE \cite{MMMR2} is a big simplification --
surprisingly this is possible without a loss of information.
However, to make it fully satisfactory, we need a maximally simple origin if this SE --
and ME would be just a dream.
Unfortunately, as anticipated in \cite{MMMish}, the story is not just so simple --
and details, though seemingly technical, are quite interesting. 
The fact that the simplest Ward identity ME is not quite the same as SE,
which controls the solution, is an interesting twist of the story
and this is what we call {\it anomaly} in the title of this paper:
\be
{\bf SE} = {\bf ME}\ {\rm mod}\ anomaly
\ee

The actual calculation consists of three steps.
First one needs to express matrix derivatives through eigenvalues,
this is discussed in a separate section \ref{eveq}).
Then one needs to act on the product ${\cal Z}_r[L] = {\cal Z}_r^{cl}[M]\cdot Z_r\{p\}$
and convert the equation w.r.t. eigenvalues into the one for the "quantum" $Z_r\{p\}$,
depending on time variables.
And afterwards one should interpret the results.
We demonstrate that "anomaly" has {\it two} origins.
The first is that
the coefficients of the terms with derivatives over $p_{kr}$
are ugly and, actually, non-polynomial in time variables.
This can serve as a possible interpretation of the need for the $r$-reduction,
i.e. the need for these derivatives to vanish --
what looks particularly convincing in the case of cubic ($r=2$) model,
when this is the only manifestation of the anomaly.
Unfortunately, for $r>2$ the situation gets more obscure.
The second phenomenon is that
for $r>2$ this non-polynomiality shows up also in the coefficients of other derivatives
-- and ME is {\it not} sufficient to explain the vanishing of these unwanted contributions.
Of course, other constituents of (\ref{wcons}) should imply this nullification,
but this brings us back to the complicated form of (\ref{wcons}) and SE (\ref{singleq}).

\section{ Strong dependence on $N$: the classical piece of partition function
\label{class}}

Usually in discussion of GKM we emphasize the remarkable property 3) from above list -- 
that the essential ("quantum") part  of partition function depends on the matrix size $N$
only through the choice of the {\it locus} $\Big\{p_k = \tr M^{-k}\Big\}$   --
an $N$-dimensional non-linear subspace
in the infinite-dimensional space of time-variables $\{p_k\}$.
The {\it shape} of $Z\{p_k\}$ is, however, independent of $N$,
and in this sense the $N$-dependence of $Z\{p\}$ on $N$ is {\it weak}.

In this paper we switch the accent to another side of the story:
to the "classical" part of partition function, which is much simpler,
but depends on $N$ much stronger -- and this will have a serious impact
on the Ward identities (\ref{wcons}) and (\ref{singleq}),
making their simplest treatment through the otherwise appealing "main equation" (\ref{maineq})
less straightforward -- if not totally meaningless.

For monomial potential $V(M) = \frac{M^{r+1}}{r+1}$
the classical part of partition function can also be easily expressed through the time variables
$p_k = \tr M^{-k}$, though expressions are a little lengthy.
They are naturally written through the Schur functions $S_R\{p\}$,
where $R$ denotes the Young diagrams (for example, $S_{[1]}=p_1,\ S_{[2]} = \frac{p_2+p_1^2}{2},\
S_{[1,1]} = \frac{-p_2+p_1^2}{2}$ an so on).
Most important, these formulas have strong and explicit dependence on $N$:
\be
 {\cal Z}^{cl}_r := \frac{e^{\tr M V'(M) - V(M)}}{\sqrt{\det V''(M)}}
 = \frac{e^{\frac{r}{r+1}\sum_{i=1}^N \mu_i^{r+1}}}
 {\prod_{i=1}^{N} \mu_i^{\frac{r-1}{2}}\prod_{i<j}^N \frac{\mu_i^r-\mu_j^r}{\mu_i-\mu_j}}
 =
\label{claZ}
\ee
\be
\!\!\!\!\!\! = \frac{S_{[1,1,\ldots,1]}^{\frac{(2N-1)(r-1)}{2}}}{S_{[(N-1)(r-1),\ldots,2(r-1),r-1]}}
 \exp\left(\frac{r}{r+1}\frac{S_{[r+1,r+1,r+1,\ldots,r+1]}-S_{[r+1,\ldots,r+1,r,1]}
 +S_{[r+1,\ldots, r+1,r,r,2]}
 - \ldots \pm S_{[r,r,\ldots,r,r,N-1]}}{S_{[1,1,\ldots,1]}^{r+1}}      \Big)   \right)
\nn
\ee
Two Young diagrams in Schur functions have $N-1$ columns,
$S_{[\underbrace{r+1,r+1,r+1,\ldots,r+1}_{N-1}]}$ in the exponent
and $S_{[\underbrace{(N-1)(r-1),\ldots,2(r-1),r-1]}_{N-1}}$ in the denominator.
All the rest have $N$ columns:
from $S_{\underbrace{[ r+1,\ldots,r+1,r,1]}_N}$
to $S_{[\underbrace{ r, \ldots,r,r,N-1}_{N}]}$ in the exponent
and also $S_{[\underbrace{1,\ldots,1}_N]} = \prod_{i=1}^N \mu_i^{-1}$.
For $N=2$ eq.(\ref{claZ}) becomes just
\be
\frac{e^{\frac{r}{r+1}( \mu_1^{r+1}+\mu_2^{r+1})}}
 {  (\mu_1\mu_2)^{\frac{r-1}{2}} \frac{\mu_1^r-\mu_1^r}{\mu_1-\mu_2}}
= \frac{S_{[1,1]}^{\frac{3(r-1)}{2}}}{S_{[r-1]}}
\exp\left(\frac{r}{r+1}\frac{S_{[r+1]}-S_{[r,1]} }{S_{[1,1]}^{r+1}}\right)
\label{claZN=2}
\ee
while for $N=3$ and $N=4$ it is
\be
\frac{e^{\frac{r}{r+1}( \mu_1^{r+1}+\mu_2^{r+1}+\mu_3^{r+1})}}
 {  (\mu_1\mu_2\mu_3)^{\frac{r-1}{2}} \frac{\mu_1^r-\mu_1^r}{\mu_1-\mu_2}
 \frac{\mu_1^r-\mu_3^r}{\mu_1-\mu_3} \frac{\mu_2^r-\mu_3^r}{\mu_2-\mu_3} }
= \frac{S_{[1,1,1]}^{\frac{5(r-1)}{2}}}{S_{[r-1,2r-2]}}
\exp\left(\frac{r}{r+1}\frac{S_{[r+1,r+1]}-S_{[r+1,r,1]}+S_{[r,r,2]}} {S_{[1,1,1]}^{r+1}}\right)
\ee
{\footnotesize
\be
\frac{e^{\frac{r}{r+1}( \mu_1^{r+1}+\mu_2^{r+1}+\mu_3^{r+1}+\mu_4^{r+1})}}
 {  (\mu_1\mu_2\mu_3\mu_4)^{\frac{r-1}{2}} \frac{\mu_1^r-\mu_1^r}{\mu_1-\mu_2}
 \frac{\mu_1^r-\mu_3^r}{\mu_1-\mu_3}\frac{\mu_1^r-\mu_4^r}{\mu_1-\mu_4} 
 \frac{\mu_2^r-\mu_3^r}{\mu_2-\mu_3} \frac{\mu_2^r-\mu_4^r}{\mu_2-\mu_4}
 \frac{\mu_3^r-\mu_4^r}{\mu_3-\mu_4} }
 = \nn 
\ee
\be
= \frac{S_{[1,1,1,1]}^{\frac{7(r-1)}{2}}}{S_{[r-1,2r-2,3r-3]}}
\exp\left(\frac{r}{r+1}\frac{S_{[r+1,r+1,r+1]}-S_{[r+1,r+1,r,1]}+S_{[r+1,r,r,2]}-S_{[r,r,r,3]}} {S_{[1,1,1,1]}^{r+1}}\right)
\nn
\ee
}

As we will see, non-trivial Schur functions in the pre-exponent
survive in the main equation (\ref{maineq})
and make it non-polynomial in time variables.
The only case when this does not matter at all, is $r=1$,
which we will briefly mention in section \ref{r1} below.
In conventional cubic model at $r=2$ the non-polynomiality can be eliminated by $r$-reduction
(from KP to KdV in this case) -- this we will see in s.\ref{r2}.
Starting from $r=3$, however, the problem (anomaly) is far more difficult to cure,
and the corrected form of {\bf ME} -- and thus the simple derivation of {\bf SE} --
still needs to be found.

\section{  From matrices to eigenvalues
\label{eveq}}

As already mentioned, GKM (\ref{GKMdef}) is an eigenvalue model,
the integral is reduced to eigenvalues of $X$ and the answer depends on the eigenvalues of $L=M^r$.
Still the reason for the special properties of GKM is that originally
it depends on the matrix variable,
and the natural Ward identities \cite{vircon})
are matrix-valued -- given by (\ref{GN}).
Since they contain matrix derivatives,
it is separate exercise to convert them to the eigenvalue form.
What we need are diagonal elements of $\frac{\p^r{\cal Z}[L]}{\p L_{tr}^r}$,
evaluated at diagonal matrix $L = {\rm diag}(\lambda_i)= {\rm diag}(\mu_i^r)$.
They do not arise from just a substitution of diagonal $L$ into ${\cal Z}[L]$.
Still the answer is well known from perturbation theory in quantum mechanics \cite{LL3}
(where one diagonalizes the Hamiltonian and obtains corrections to the wave functions):
according to  \cite{MMM},
\be
\left(\frac{\p^r F}{\p L_{tr}^r}\right)_{ii} = \sum_{j_1,\ldots,j_{r-1}}
\left(
\sum_{{\rm permutations\ of}\ i,j_1,\ldots,j_{r-1}}
\frac{\frac{\p F}{\p \lambda_i}}{\prod_{s=1}^{r-1}(\lambda_i - \lambda_{j_s})}
\right)
\ee
Note that $\lambda_{j_s}$ can coincide, also with $\lambda_i$ --
then one should apply the l'Hopitale rule,
and this gives rise to more sophisticated structures.
In \cite{Mikh}
a special technique was developed on this occasion.
We, however, just work with explicit formulas,
without going into details of the derivations.
In particular,
\be
\left(\frac{\p  F}{\p L_{tr} }\right)_{ii} = \frac{\p F}{\p \lambda_i},
\nn \\
\left(\frac{\p^2 F}{\p L_{tr}^2 }\right)_{ii}
=\sum_j \frac{\frac{\p F}{\p \lambda_i}-\frac{\p F}{\p \lambda_j}}{\lambda_i-\lambda_j}
=  \sum_{j\neq i} \frac{\frac{\p F}{\p \lambda_i}-\frac{\p F}{\p \lambda_j}}{\lambda_i-\lambda_j}
+ \frac{\p^2 F}{\p \lambda_i^2},
\nn \\
\left(\frac{\p^3 F}{\p L_{tr}^3 }\right)_{ii} =
\sum_{j,k} \frac{\frac{\p F}{\p \lambda_i} }{(\lambda_i-\lambda_j)(\lambda_i-\lambda_k)}
+ \frac{\frac{\p F}{\p \lambda_j} }{(\lambda_j-\lambda_i)(\lambda_j-\lambda_k)}
+ \frac{\frac{\p F}{\p \lambda_k} }{(\lambda_k-\lambda_i)(\lambda_k-\lambda_j)}
= \nn \\
= \sum_{k\neq j\neq i}^N \left(
\frac{\frac{\p F}{\p \lambda_i}}{(\lambda_i-\lambda_j)(\lambda_i-\lambda_k)}
+ \frac{ \frac{\p F}{\p \lambda_j}}{(\lambda_j-\lambda_i)(\lambda_j-\lambda_k)}
+ \frac{ \frac{\p F}{\p \lambda_k}}{(\lambda_k-\lambda_i)(\lambda_k-\lambda_j)}\right)
- \sum_{j\neq i}^N  \frac{\frac{\p F}{\p \lambda_i}
-  \frac{\p F}{\p \lambda_j}}{(\lambda_i-\lambda_j)^2}
+ \nn \\
+\sum_{j\neq i}^N
\frac{  2 \frac{\p^2 F}{\p \lambda_i^2}
-   \frac{\p^2 F}{\p \lambda_i\p\lambda_j} -  \frac{\p^2 F}{\p \lambda_j^2}}
{\lambda_i-\lambda_j}
+ \frac{\p^3 F}{\p \lambda_i^3},
\nn \\
\ldots
\label{evders}
\ee
For example, at $N=2$, for a function {\footnotesize $F(\lambda_1,\lambda_2):=\!\!
F\!\left(\!\frac{L_{11}+L_{22}\pm\sqrt{(L_{11}-L_{22})^2+4L_{12}L_{21}}}{2}\!\right)$}
one can explicitly check, that
\be
\!\!\!\!\!
\left\{\left.\left(\frac{\p^3 F}{\p \L_{\rm tr}^3}\right)_{11}\!\!\!\! :=
\frac{\p^3 F}{\p \L_{11}^3} + 2\frac{\p^3 F}{\p \L_{11}\p L_{12} \p L_{21}}
+ \frac{\p^3 F}{\p \L_{12}\p L_{21} \p L_{22}}\right\}\right|_{L={\rm diag}(\lambda_1,\lambda_2)}
\!\!\!\!\! = -\frac{F_{,1}-F_{,2}}{(\lambda_1-\lambda_2)^2}
+ \frac{2F_{,11} - F_{,12}-F_{,22}}{\lambda_1-\lambda_2}  + F_{1,1,1}
\nn
\ee
in accordance with this general prescription.

\section{A toy example at $r=1$
\label{r1}}

This is a special case, where equation (\ref{GN}) has a "wrong" power of $\p/\p L$.
It was used as a training  example in \cite{MMM}.
For $Z\{p_k\} = Z\left\{\sum_{i=1}^N \lambda_i^{-k}\right\}$ we have:
\be
\sum_{i=1}^N \left(\frac{\partial^2 {\cal Z}}{\partial L_{tr}^2}\right)_{ii} =
\sum_{i\neq j}^N \frac{\frac{\p {\cal Z}}{\p \lambda_i}
-  \frac{\p {\cal Z}}{\p \lambda_j}}{\lambda_i-\lambda_j}
+ \sum_{i=1}^N \frac{\p^2 {\cal Z}}{\p\lambda_i^2}
= \sum_{n=1}^\infty \sum_{a=1}^{n+1} n p_a p_{n+2-a} \frac{\p Z}{\p p_n}
+ \sum_{n_1,n_2=1}^\infty n_1n_2 p_{n_1+n_2+2}\,\frac{\p^2Z}{\p p_{n_1}\p p_{n_2}}
= \nn \\
= \sum_{n=-1}^{\infty} p_{n+2} \left(\sum  (k+n)p_{k} \frac{\p Z}{\p p_{k+n}}
+ \sum_{a+b=n} ab \frac{\p^2 Z}{\p p_ap_b}\right)
= \sum_{n=1}^{\infty} p_{n} \hat L_{n-2}^{(1)} Z
\ee
where
\be
\hat L_n^{(1)} = \sum_{k=1}^\infty  (k+n)p_{k} \frac{\p Z}{\p p_{k+n}} + \sum_{a+b=n} ab \frac{\p^2 Z}{\p p_ap_b}
\ee
and superscript label $(1)$ refers to $r=1$.
This is the ordinary Virasoro operator, which defines Virasoro constraints in Hermitian matrix model,
and it appears here because this model can be also treated as GKM with additional insertion
of a power of $\det X$ in the integral, what causes also an increase of $r$ by one
\cite{CheMa,UFN3}.

\section{Original cubic ($r=2$) Kontsevich model  \label{r2}}

\subsection{Implication from the known $Z_2$}

We now proceed to the study of the true main equation (\ref{maineq}),
beginning from the first case of cubic Kontsevich model.
What we need is to substitute
\be
{\cal Z}_2 = \overbrace{\frac{e^{\frac{2}{3}\sum \mu_i^{3}}}{\sqrt{\prod_i {\mu_i}}\prod_{i<j}(\mu_i+\mu_j)}}
^{{\cal Z}_2^{cl}}
\cdot Z_2\{p_k\}
\label{Z2}
\ee
into (\ref{maineq}):
\be
\frac{1}{{\cal Z}_2^{cl}}\sum_{i=1}^N  \sqrt{\lambda_i}\Big(
\overbrace{\frac{\p^2 {\cal Z}_2}{\p\lambda_i^2}
+ \sum_{j\neq i}^N \frac{\frac{\p {\cal Z}_2}{\p \lambda_i}
-  \frac{\p {\cal Z}_2}{\p \lambda_j}}{\lambda_i-\lambda_j}
}^{(\partial^2 {\cal Z}_2/{\partial L_{tr}^2})_{ii}}
- \lambda_i {\cal Z}_2 \Big)
= \sum_{i=1}^N  \sqrt{\lambda_i}\left(
\frac{\p^2 {  Z}_2}{\p\lambda_i^2}
+\sum_{j\neq i}^N \frac{\frac{\p {  Z}_2}{\p \lambda_i}
-  \frac{\p {  Z}_2}{\p \lambda_j}}{\lambda_i-\lambda_j}
 \right) 
+ \nn
\ee
\be
+  \sum_{i=1}^N  \sqrt{\lambda_i}\left(
  \frac{\p^2 \log{\cal Z}_2^{cl}}{\p\lambda_i^2}
  + \left(\frac{\p  \log{\cal Z}_2^{cl}}{\p\lambda_i }\right)^2
  + \sum_{j\neq i}\frac{\frac{\p \log{\cal Z}_2^{cl}}{\p \lambda_i}
-  \frac{\p \log{\cal Z}_2^{cl}}{\p \lambda_j}}{\lambda_i-\lambda_j} - \lambda_i\right)Z_2
+   2\sum_{i=1}^N  \sqrt{\lambda_i} \frac{\p \log{\cal Z}_2^{cl}}{\p\lambda_i}
\frac{\p Z_2}{\p\lambda_i}
\label{ME2}
\ee
First of all we can substitute the known series for $Z_2\{p_k\}$
(last time cited in the Appendix to \cite{MMMish}),
\be
Z_2\{p_k\} = 1 + \left(\frac{p_2p_1^2}{6}+ \frac{p_4}{36}\right)
+ \left( \frac{13}{216}p_4p_2p_1^2+\frac{13}{2592}p_4^2 - \frac{1}{216}p_2^4
+ \frac{1}{72}p_2^2p_1^4 + \frac{1}{27}p_5p_1^3 + \frac{1}{27}p_7p_1   \right) + \ldots
\ee
and we expect to get zero.
It is instructive to see how this really works.
If we substitute instead of $Z_2$ just $1$ -- the first term in the series,-- 
we get a polynomial of grading 3:  $\frac{p_3+4p_1^3}{16}$,
if $1 + \left(\frac{p_2p_1^2}{6}+ \frac{p_4}{36}\right)$, then a polynomial of grading 6
and so on: the more gradings we include into $Z_2\{p\}$, the higher is the grading of (\ref{ME2}):
if gradings up to $3m$ are included into $Z_2\{p\}$, then (\ref{ME2}) is of grading $3(m+1)$.
Thus we obtain zero for (\ref{ME2}) in the sense that every particular grading vanishes
is we include appropriately many terms into $Z_2\{p\}$.
Also at every stage the answer is not just of definite grading,
it is actually a {\it polynomial} in time variables.

One can wonder, what happens to $S_{[1]}$ factor in (\ref{claZN=2}) -- 
why does not it produce non-polynomial contributions?
It turns out to be a rather delicate adjustment.
Already the quadratic singularity $S_{[1]}^{-2} = (\mu_1+\mu_2)^{-2}$ drops out from the sum
$\sum_i \mu_i \frac{\p^2 (\mu_1+\mu_2)^\alpha}{\p \lambda_i^2}$
because of the peculiar property $\frac{\mu_1}{\mu_1^2} + \frac{\mu_2}{\mu_2^2} \sim \mu_1+\mu_2$.
The linear singularity is even more miraculous: 
potentially relevant terms in (\ref{claZN=2}) are
\be
\sum_i \mu_i \left\{\frac{\p^2 \log \frac{1}{S_{[1]}}}{\p \lambda_i^2} 
+ \left(  
\frac{    \p  \left( \log \frac{S_{[1,1]}^{\frac{3\beta}{2}}}{S_{[1]}} 
+ \frac{2\alpha(S_{[3]}-S_{[2,1]})}{3S_{[1,1]}^3}\right) }
{\p \lambda_i}
\right)^2
+ \gamma\sum_{j\neq i} \frac{\frac{\p \log \frac{1}{S_{[1]}}}{\p \lambda_i}-
\frac{\p \log \frac{1}{S_{[1]}}}{\p \lambda_j}}{\lambda_i-\lambda_j}\right\}
\ee
and the term $\frac{1}{\mu_1+\mu_2}$ is independent of $\alpha$, but depends in $\beta$.
It vanishes when $\gamma=3\beta-2$, what includes the true values $\beta=\gamma=1$,
but clearly demonstrates the delicate balance between various contributions.
Therefore it is not surprising that thus balance will be often violated --
the surprise is that it continues to hold (anomaly is lacking) for $r=2$ for arbitrary $N$,
and also for the coefficients of odd derivatives $\frac{\p Z}{\p p_{2n+1}}$ --
as we will explain in the next sections.

\subsection{$r=2$, all times}

The next exercise is to convert (\ref{ME2}) into an equation for $Z_2\{p_k\}$,
similar to what we considered above in s.\ref{r1}.
Assume first that $Z_2\{p_k\}$ depends on all the time-variables $p_k$.
Then, once again substituting (\ref{Z2}) into (\ref{ME2}) we get:
\be
0 \ \stackrel{(\ref{maineq})}{=} \
\frac{1}{{\cal Z}_2^{cl}}\sum_{i=1}^N  \sqrt{\lambda_i}\Big(
\overbrace{ \frac{\p^2 {\cal Z}_2}{\p\lambda_i^2}
+ \sum_{j\neq i}^N \frac{\frac{\p {\cal Z}_2}{\p \lambda_i}
-  \frac{\p {\cal Z}_2}{\p \lambda_j}}{\lambda_i-\lambda_j}
}^{(\partial^2 {\cal Z}_2/{\partial L_{tr}^2})_{ii}}
- \lambda_i {\cal Z}_2 \Big)=
\label{ME2alltimes}
\ee
\be
\!\!\!\!\!\!\!\!
= \frac{1}{4}\sum_{n_1,n_2=1}^\infty n_1n_2 p_{n_1+n_2+3}\,\frac{\p^2Z}{\p p_{n_1}\p p_{n_2}}
+ \sum_{n=1}^\infty n\left(-p_n + \frac{1}{2}\sum_{a=1}^{\frac{n-1}{2}+2}   p_{2a-1} p_{n+4-2a}
+\underline{\xi_n^{(N)}\cdot  \frac{{\rm frac}\left(\frac{n-1}{2} \right)}{2}}
\right) \frac{\p Z}{\p p_n}
+  \frac{p_3+4p_1^3}{16} \, Z
\nn
\ee
The last term is of course the same as we got from substitution of $1$ into (\ref{ME2}).
However, with the first derivatives of $Z$ there is a trouble (underlined):
for derivatives w.r.t. even $p_{2k}$ the coefficients are not polynomial in $p$.
Moreover, they depend on $N$, with somewhat sophisticated self-consistency/reduction relations
between different $N$.
In terms of Schur functions $S_R$
\be
\xi_n^{(2)} =    \frac{p_{n+4}+(p_1^2-p_2)p_{n+2}}{p_1} = \frac{S_{[n+4]} + 2S_{[n+3,1]}-3S_{[n+2,2]}}{S_{[1]}}
\nn \\
\xi_n^{(3)} = \frac{S_{[n+5,1]}+2S_{[n+4,2]}+2S_{[n+4,1,1]}-3S_{[n+3,3]} -
 S_{[n+2,3,1]}-2S_{[n+2,2,2]}+5S_{[n+1,3,2]}}{2S_{[2,1]}}
\nn
\ee
{\footnotesize
\be
\!\!\!\!\!\!\!\!
\xi_n^{(4)} = \frac{S_{[n+6,2,1]}+2S_{[n+5,3,1]}+2S_{[n+5,2,2]}+2S_{[n+5,2,1,1]}-3S_{[n+4,4,1]}
-2S_{[n+3,4,2]}-2S_{[n+3,3,3]}
 - 2S_{[n+3,2,2,2]} - 2S_{[n+3,4,1,1]}+5S_{[n+2,4,3]}
}{S_{[3,2,1]}}
+ \nn \\
+ \frac{1}{S_{[3,2,1]}}\cdot\left\{ \begin{array}{ccc}  9S_{[n+1,3,3,2]} &{\rm for} & n=2 \\ \\
2S_{[n+1,3,3,2]} + 2S_{[n+1,4,3,1]} + 2S_{[n+1,4,2,2]} - 7S_{[n,4,3,2]} &{\rm for} & n\geq 4
\end{array}\right.
\nn
\ee
}
\be
\ldots
\ee
i.e. denominator is equal to $S_{[N-1,\ldots,3,2,1]}$.

This problem of non-polynomiality  is cured (the underlined terms are absent)
in the action on functions $Z_2\{p_1,p_3,p_5,\ldots\}$,
which depend only on odd times $p_{2k-1}$.

\subsection{$r=2$, odd times = cubic Kontsevich model
\label{r2odd}}

Assume now that $Z\{p_k\}$ depends on all the odd time-variables $p_{2k-1}$.
Then the terms with $\xi^{(N)}$ drop out of (\ref{ME2alltimes}) and we get a
differential  equation for $Z-2\{p\}$ with  polynomial coefficients in $p$:
\be
0 \ \stackrel{(\ref{maineq})}{=} \
\sum_{i=1}^N  \lambda_i^{1/r}\Big(
\overbrace{\sum_{j\neq i}^N \frac{\frac{\p Z}{\p \lambda_i} -  \frac{\p Z}{\p \lambda_j}}{\lambda_i-\lambda_j}
+  \frac{\p^2 Z}{\p\lambda_i^2}}^{(\partial^2 {\cal Z}/{\partial L_{tr}^2})_{ii}} - \lambda_i Z \Big)
= \nn
\ee
\be
\!\!\!\!\!\!\!\!\!\!\!\!\!\!\!\!\!\!\!\!\!\!\!\!\!\!\!
=
\frac{1}{4}\sum_{n_1,n_2=1}^\infty (2n_1-1)(2n_2-1) p_{2n_1+2n_2+1}\,\frac{\p^2Z}{\p p_{2n_1-1}\p p_{2n_2-1}}
+  \sum_{n=1}^\infty (2n-1)\left(-p_{2n-1}
+\frac{1}{2}\sum_{a=1}^{n}  p_{2a-1} p_{2n+3-2a}\right) \frac{\p Z}{\p p_{2n-1}}
+  \frac{p_3+4p_1^3}{16} \, Z
= \nn
\ee
\be
= -\hat l_0 Z + \sum_{n=1}^{\infty} p_{2n-1} \hat L_{n-2}^{(2)} Z
\label{ME2}
\ee
where
\be
\hat L_n^{(2)} = \frac{1}{4}\delta_{n,0} + \frac{p_1^2}{16}\delta_{n,-1}
+\frac{1}{2}\sum_{k=1}^\infty  (2k+2n-1)p_{2k-1} \frac{\p  }{\p p_{2k+2n-1}} +
\frac{1}{4} \sum_{a+b=2n} (2a-1)(2b-1) \frac{\p^2  }{\p p_{2a-1}\p p_{2b-1}}
\ee
and the grading-counting operator
\be
\hat l_0 = \sum_{n=1}^\infty (2n-1)p_{2n-1} \frac{\p}{\p p_{2n-1}}
\ee

Note that elimination of derivatives $\frac{\p Z}{\p p_{2k}}$ automatically
eliminates all even times $p_{2k}$ from the coefficients of (\ref{ME2}):
{\bf for $r=2$ the $r$-reduction is necessary and sufficient for {\bf ME} to reproduce {\bf SE}}.

\section{The first non-standard case: quartic model $r+1=4$}

\subsection{Solution to projected Ward identity}

Now we can repeat all the same steps in the first non-trivial case of quartic GKM with $r=3$.
We will see that the non-polynomiality gets now even more pronounced.

In terms of $\mu$-variables ($\lambda_i = \mu_i^r$)
the {\it main equation} in this case looks as follows:
\be
\!\!\!\!\!\!\!\!\!\!\!\!\!\!\!\!\!\!\!\!\!\!\!\!\!\!\!\!
\sum_{i=1}^N \lambda_i^{1/3}\left\{
\sum_{k\neq j\neq i}^N \left(
\frac{\frac{\p{\cal Z}}{\p \lambda_i}}{(\lambda_i-\lambda_j)(\lambda_i-\lambda_k)}
+ \frac{ \frac{\p{\cal Z}}{\p \lambda_j}}{(\lambda_j-\lambda_i)(\lambda_j-\lambda_k)}
+ \frac{ \frac{\p{\cal Z}}{\p \lambda_k}}{(\lambda_k-\lambda_i)(\lambda_k-\lambda_j)}\right)
+ \right.\nn \\ \left.
+\sum_{j\neq i}^N \left(-\frac{   \frac{\p{\cal Z}}{\p \lambda_i}
-  \frac{\p{\cal Z}}{\p \lambda_j}}{(\lambda_i-\lambda_j)^2}
+ \frac{  2 \frac{\p^2{\cal Z}}{\p \lambda_i^2}
-   \frac{\p^2{\cal Z}}{\p \lambda_i\p\lambda_j} -  \frac{\p^2{\cal Z}}{\p \lambda_j^2}}
{\lambda_i-\lambda_j}
\right)
+ \frac{\p^3 {\cal Z}}{\p \lambda_i^3} - \lambda_i {\cal Z}
\right\} \ \stackrel{(\ref{maineq})}{=}\ 0
\label{ME3ev}
\ee
Conversion to $\mu$-variables ($\lambda = \mu_i^r$) is easy:
$\frac{\p}{\p\lambda_i} = \frac{1}{r\mu_j^{r-1}} \frac{\p}{\p\mu_i}$.
For $r>3$ denominators get larger and degenerations provide higher derivatives of ${\cal Z}$.

The next step is to substitute
\be
{\cal Z}_3 = \frac{e^{\frac{3}{4}\tr M^{4}}}
{\sqrt{\det \left(M^2\otimes 1 + M\otimes M + 1\otimes M^2\right)}}\cdot Z_3\{p_k\}
= \frac{e^{\frac{3}{4}\sum_i \mu_i^4}}{\prod_{i} \mu_i \prod_{i<j} (\mu_i^2+\mu_i\mu_j+\mu_j^2)}
\cdot Z_3\{p_k\}
\ee
and obtain an equation ME (\ref{maineq}) for $Z_3$ with time-derivatives instead of the $\mu$-ones.

Then we can compare it to SE (\ref{singleq}), which in the case of $r=3$ involves two operators \cite{MMMish}:
\be
\hat W^{(2)}_n = \frac{1}{3}\sum_{k=1}^\infty (k+3n)P_k\frac{\p}{\p p_{k+3n}}
+ \frac{1}{6} \sum_{a+b=3n} ab\frac{\p^2}{\p p_a\p p_b}
+ \frac{p_1p_2}{3}\delta_{n,-1} + \frac{1}{9}\delta_{n,0}
\ee
{\footnotesize
\be
\hat W^{(3)}_n = \frac{1}{9}\sum_{k,l=1}^\infty (k+l+3n)P_kP_l\frac{\p}{\p p_{k+l+3n}}
+ \frac{1}{9}\sum_{k=1}^\infty \sum_{a+b=k+3n} abP_k\frac{\p^2}{\p p_a\p p_b}
+ \frac{1}{27}\!\!\sum_{a+b+c=3n}abc\frac{\p^3}{\p p_a\p p_b\p p_c}
+ \frac{1}{27}\!\!\sum_{a+b+c=-3n} P_aP_bP_c
\nn
\ee
}

\noindent
with $P_k=p_k - 3\delta_{k,4}$ and $a,b,c,k,l$ not divisible by $3$.
Note that the sums are restricted more than it would follow from omission of derivatives
w.r.t. $p_{3k}$, for example, there are no terms $\frac{\p^3}{\p p_1^2\p p_2}$
and $\frac{\p^3}{\p p_1\p p_2^2}$ in $\hat W^{(3)}$, only $\frac{\p^3}{\p p_1^3}$
and $\frac{\p^3}{\p p_2^3}$.
This will be one of the apparent differences from the ME,
which contains third derivatives of all the four kinds.

\subsection{Non-trivial denominators}

The other striking difference will be non-polynomiality.
In fact, one can observe it at the very early stage.
For $r>2$ it is enough to look at the derivative-free term in ME.
Namely, if $Z_3=1$ the l.h.s. of (\ref{ME3ev}) is non-vanishing, but
contains  contributions of just two ($r-1$) gradings:  $-4$ and $0$.
In the simplest case of $N=2$
\be
Z = 1 \ \ \ \stackrel{(\ref{ME3ev})}{\Longrightarrow} \ \ \
\frac{7S_{[4]}+5S_{[3,1]}}{9} \underline{-
\frac{4\,\big(7S_{[10]}+7S_{[9,1]}+10S_{[8,2]}\big)}{27S_{[2]}}
}
\label{Z3=1}
\ee
The first is polynomial in times, the second is not.
In other words we observe the same phenomenon as in (\ref{ME2alltimes}),
but now it is present already for the item $Z$, without derivatives.

Adding appropriate $p$-dependent pieces to $Z_3\{p\}$ \cite{MMMish}
preserves the pattern --
just shifts it to higher and higher gradings:
{\footnotesize
\be
\!\!\!\!\!\!\!\!\!\!\!\!
Z = 1 + \left(\frac{p_4}{36} + \frac{p_2p_1^2}{6}\right) \ \ \ \Longrightarrow \ \ \
\frac{35p_4\big(11S_{[4]} +13S_{[3,1]}\big)}{324} - \frac{35\big(32S_{[13,1]}+38S_{[12,2]}
+ S_{[2]}(22S_{[12]}-22S_{[11,1]}+35S_{[9,3]}-36S_{[8,4]}+35S_{[6,6]}) \big)}{243 S_{[2]}}
\nn
\ee
}

\noindent
and so on.
For generic $N$ denominator becomes $S_{[2N-2,\ldots,6,4,2]}$:
\be
Z_3 = 1 \ \ \ \stackrel{(\ref{ME3ev})}{\Longrightarrow} \ \ \
\overbrace{\frac{7S_{[4]}+5S_{[3,1]}-5S_{[2,1,1]} - 7 S_{[1,1,1,1]}}{9}}^{\frac{p_4+6p_2p_1^2}{9}}  -
\label{Z3=1N}
\ee
{\footnotesize
\be
\!\!\!\!\!\!
-  \left\{ \begin{array}{lcc}
\frac{4\,\Big(7S_{[10]}+7S_{[9,1]}+10S_{[8,2]}\Big)}{27S_{[2]}}
= \frac{1}{27}\left((p_2+p_1^2)(4p_2^3+21p_2^2p_1^2-12p_2p_1^4+p_1^6)
 -\frac{12p[2]S_{[1,1]}^4}{S_{[2]}}\right)
 & & N=2 \\
\frac{1}{27S_{[4,2]}}\Big(28S_{[12,2]}+28S_{[11,3]}+28S_{[11,2,1]}+40S_{[10,4]}+18S_{[10,3,1]}
+ 40S_{[10,2,2]} + 30S_{[9,4,1]} + 30S_{[9,3,2]}
-\nn \\
- 16S_{[8,5,1]}+42S_{[8,4,2]} -16S_{[8,3,3]}+3S_{[7,6,1]}
-13S_{[7,5,2]} - 22S_{[7,4,3]}+3S_{[6,6,2]}-19S_{[6,5,3]}-6S_{[6,4,4]}+12S_{[5,5,4]}   \Big) & & N=3 \\
\frac{1}{27S_{[6,4,2]}}\Big(\ldots  \Big) & & N=4 \\
\ldots
\end{array}\right.
\ee
}

\noindent
While the first polynomial piece is stabilized and does not vary anymore for $N>r$,
the shape of non-polynomial terms is not stable -- it varies with $N$.

Building up the true $Z_3\{p\}$ results into the shift of the two non-vanishing
pieces to infinite gradings $p_\infty$, $p^\infty$ --
and in this sense the answer, understood as the contributions at every particular grading,
gets vanishing.

The   moral is that now the non-polynomiality is less related to $r$-reduction:
a function can be independent of $p_{rk}$ (like $Z_3=1$), still (\ref{ME3ev})
does {\it  not} convert it into a {\it polynomial} -- non-trivial denominators occur.
However, the {\it proper} $Z_3\{p\}$ is converted to zero.
Together with occurrence of the underlined term in (\ref{Z3=1})
this implies that at least some terms in the {\it main} equation
with derivatives of $Z_3\{p_k\}$ w.r.t. $p_k$ should be non-polynomial,
even if $k$ is {\it not} divisible by $r$.
Since such non-polynomiality does not appear in the highest-derivative terms $\frac{\p^{r}Z_r}{\p p_{i_1}\ldots \p p_{i_{r}}}$,
the natural guess after that is that these additional terms are made from the
lower $W$-constraints, i.e. from the complements of the main equation (\ref{maineq})
-- the other corollaries of the matrix Ward identity (\ref{GN}).

\subsection{ME for $r=3$, all times}

As we already know from the previous subsection, there will be problems with relating SE to ME.
In addition to the two nice terms at the r.h.s. of
{\footnotesize
\be
\!\!\!\!\!\!\!\!\!\!\!\!\!\!\!\!\!\!\!\!\!\!\!\!\!\!\!
0 \ \stackrel{(\ref{maineq})}{=} \
\sum_{i=1}^N  \lambda_i^{1/r}\Big(
\overbrace{  \sum_{k\neq j \neq i}^N
\frac{\frac{\p Z}{\p \lambda_i} }{(\lambda_i-\lambda_j)(\lambda_i-\lambda_k)}
- \sum_{j \neq i}^N  \frac{\frac{\p^2 Z}{\p\lambda_i^2} }{\lambda_i-\lambda_j}
+  \frac{\p^3  Z}{\p\lambda_i^3 }
}^{(\partial^3 {\cal Z}/{\partial L_{tr}^3})_{ii}}\  -\  \lambda_i Z \Big)
= \nn
\ee
}
\be
= \frac{1}{27}\sum_{n_1,n_2,n_3=1}^{\infty} n_1n_2n_3 p_{n_1+n_2+n_3+8}\,\frac{\p^3Z}{\p p_{n_1}\p p_{n_2}\p p_{n_3}}
+\frac{1}{3}\sum_{n_1,n_2 }^{\infty} n_1n_2  p_{n_1+n_2+4}\,\frac{\p^2Z}{\p p_{n_1}\p p_{n_2} }
+ \ldots
\label{ME3}
\ee
the non-polynomial terms will appear,
which depend on $N$.
In the simplest case of $N=2$ the full expression is:

{\footnotesize
\be
0 \ \stackrel{(\ref{maineq})}{=} \
-\!\!\!\!\!\sum_{n_1,n_2,n_3=1}^\infty\!\!\!\!\!\!\!\!
\frac{n_1n_2n_3p_{n_1+n_2+n_3+8}}{27}\frac{\p^3 Z}{\p p_{n_1}\p p_{n_2}\p p_{n_3}}
+ \nn \\
+ \sum_{n_1,n_2 =1}^\infty  n_1n_2\left(\frac{p_{n_1+n_2+4}}{3} - \frac{(n_1+n_2+8)p_{n_1+n_2+8}}{18}
- \frac{ 2p_{n_1+n_2+6}S_{[1,1]} + 4p_{n_1+n_2+4}S_{[2,2]} + 2p_{n_1+n_2}S_{[4,4]}
+4p_{n_1+n_2-2}S_{[5,5]}}{18}
+ \right.\nn \\ \left. \!\!\!\!\!\!\!\!\!\!\!\!\!\!\!\!\!\!\!\!\!\!\!\!\!\!\!\!\!\!\!\!\!\!\!\!\!\!\!\!
+ \frac{S_{[1,1]}^7\cdot \big(S_{[n_1+n_2-4]} + 2S_{[n_1+n_2-5,1]}+S_{[n_1-2,n_2-2]}\big)
-\frac{S([1,1])^{n_2+5}}{2}p_1S_{[n_1-n_2-1]} + S_{[1,1]}^6S_{[n_1-1]}\delta_{n_2,1}}{9S_{[2]}}
-\right. \nn \\ \left.
-\frac{ 2S_{[1,1]}^7\cdot\big(\delta_{n_1,3}\delta_{n_2,1}+\delta_{n_1,2}\delta_{n_2,2}  \big)
+S_{[1,1]}^6\big(2p_1\delta_{n_1,2}+\delta_{n_1,1}   \big)\delta_{n_2,1} }{9S_{[2]}}
\right)\frac{\p^2 Z}{\p p_{n_1}\p p_{n_2} }
+ \nn \\   \!\!\!\!\!\!\!\!\!\!\!\!\!\!\!\!\!\!\!\!\!\!\!\!\!\!\!\!\!\!\!\!\!\!\!\!\!\!\!\!\!\!\!\!
+\sum_n n\left(
\frac{(n+4)\big(S_{[n+6]} -S_{[n+3,2]}\big)+ 3\big(S_{[n+5,1]}+S_{[n+4,2]}\big)}{3S_{[2]}}
-\right. \nn \\ \left.
- \frac{(n^2+12n+39)S_{[n+10]} + 3(n+7)S_{[n+9,1]}+6(n+6)S_{[n+8,2]} - (n^2+12n+27)S_{[n+7,3]}
- 3(1-\delta_{n,1})S_{[n+4,6]}  }{27S_{[2]}}
\right) \frac{\p Z}{\p p_n}
- \nn \\
-np_n \frac{\p Z}{\p p_n}
+ \left(\frac{7S_{[4]}+5S_{[3,1]}}{9} - \frac{4\big(7S_{[10]}+7S_{[9,1]}+10S_{[8,2]}\big)}{27S_{[2]}}\right) Z
\nn
\ee
}

\subsection{ME versus SE}

This should be compared to the operator in (\ref{singleq}), which after substitution of (\ref{wcons})
becomes
\be
- \!\!\!\!\!\!\! \sum_{a+b+c =3n-9} \!\!\!\!\frac{abcp_{a+b+c+8}}{27}\frac{\p^3 Z}{\p p_a\p p_b\p p_c}
+ \left(\frac{1}{3}\sum_{a+b=3n-5}+\frac{1}{6}\sum_{a+b=3n-6}\right) ab p_{a+b+4} \frac{\p^2 Z}{\p p_a\p p_b} - \sum_{a+b=3n-1}\!\!\!\!\frac{ab p_k p_{a+b-k+8}}{9}\frac{\p^2 Z}{\p p_a\p p_b}
+ \nn \\
 + \frac{2}{3}p_{3n-1}(k+3n-5)p_k \frac{\p Z}{\p p_{k+3n-5}}
+ \frac{1}{3}p_{3n-2}(k+3n-6)p_k\frac{\p Z}{\p p_{k+3n-6}}
- \frac{1}{9}p_{3n-1}(k+l+3n-9)p_kp_l \frac{\p Z}{\p p_{k+l+3n-9}}
- \nn
\ee
\be
-np_n \frac{\p Z}{\p p_n}
+  \left(\frac{p_4+6p_2p_1^2}{9} - \frac{p_5p_1^3+3p_4p_2p_1^2+p_2^4}{27}\right)Z
\ \stackrel{(\ref{singleq})}{=} \ 0
\label{SE3}
\ee

Like it was for $r=2$,
in (\ref{ME3}) there are items with the derivatives $\p/\p p_{3k}$,
which are absent in (\ref{SE3}).
They can be eliminated by asking $Z_3$ to belong to the $3$-reduction --
exactly like it happened in the previous section for $r=2$.
This is the positive part of the story:
{\bf cancellation of anomaly requires the $r$-reduction}.
But is the $r$-reduction sufficient for deriving {\bf SE} from {\bf ME}?

Unfortunately, the answer is "no": now there are a few more striking differences
between (\ref{ME3}) and (\ref{SE3}), e.g.

i) already in (\ref{ME3}) there are items with $p_{3k}$ in the sums,
which are absent in (\ref{SE3}),

ii) the full expression at the r.h.s. of (\ref{ME3}) contains non-polynomial
terms with denominators $S_{[\ldots,2]}$,

iii) terms like $\p^3_{112} Z$, $\p^3_{122} Z$ and $\p^2_{11} Z$, $\p^2_{22}Z$
are present already in (\ref{ME3}),
while only $\p^3_{111}Z$, $\p^3_{222}Z$ and $\p^2_{12}Z$ are allowed in (\ref{SE3}).

These are the {\it qualitative} deviations, as to the quantitative {\it details} of the two formulas,
they look  even more different.
Still both are true.
The only way out of this apparent discrepancy is that the {\it anomalous} difference
between the two formulas is made from some other $W$-constraints (\ref{wcons}),
not incorporated into the simple equation (\ref{SE3}).
This would mean that literally ${\bf SE}\neq {\bf ME}$ even for $r$-reduced $Z_r$,
still the anomaly is canceled by $r$-reduction {\it plus} some additional information --
superficial for the scalar projection {\bf ME} of the Ward identity
(\ref{GN}), still implied by the entire (\ref{GN}).
This is indeed the case, but it is quite difficult to see.
We show how it works for contributions from a few lowest  gradings $4m$ to
$Z_4 = \sum_{m=0}^\infty z_{4m}$.

In grading four we have exact matching: both (\ref{ME3}) and (\ref{SE3}) contribute
\be
\underline{-np_n \frac{\p z_4}{\p p_n}}
+  \frac{p_4+6p_2p_1^2}{9} z_0 = \left(-\frac{4}{4}+1\right)\cdot \frac{p_4+6p_2p_1^2}{9} = 0
\ee

In grading eight:
\be
{\bf SE}: \ \ \ \ \ \ \ \ \ \ \ \
\underline{-np_n \frac{\p z_8}{\p p_n}}
+  \frac{p_4+6p_2p_1^2}{9} z_4
- \frac{p_5p_1^3+3p_4p_2p_1^2+p_2^4}{27} z_0
+ \ \ \ \ \ \ \ \ \ \ \nn \\
+ \sum_n\sum_{k\neq 0 \ {\rm mod}\ 3}\left(\frac{2}{3}p_{3n-1}(k+3n-5)p_k \frac{\p z_4}{\p p_{k+3n-5}}
+ \frac{1}{3}p_{3n-2}(k+3n-6)p_k\frac{\p z_4}{\p p_{k+3n-6}}\right)
+\nn \\
+  \Big(\!\!\!\!\!\!\underbrace{\frac{1}{3}\sum_{a+b=3n-5}}_{\text{does not contribute}}
+\frac{1}{6}\sum_{a+b=3n-6}\Big)\ ab p_{a+b+4} \frac{\p^2 z_4}{\p p_a\p p_b}
=  0
\label{SE31}
\ee
while (for $N=2$)
\be
{\bf ME}: \ \ \ \ \ \ \ \ \ \ \ \
\underline{-np_n \frac{\p z_8}{\p p_n}}  +  \frac{p_4+6p_2p_1^2}{9} z_4
- \frac{4(7S_{[10]}+7S_{[9,1]}+10S_{[8,2]})} {27S_{[2]}} z_0
+ \ \ \ \ \ \ \ \ \ \ \\
+\frac{1}{S_{[2]}}\sum_n  n\left(\frac{n+4}{3}(S_{[n+6]}-S_{[n+3,3]})+(S_{[n+5,1]}+S_{[n+4,2]})\right)
\frac{\p z_4}{\p p_n}
+\frac{1}{3}\sum_{n_1,n_2} n_1n_2 p_{n_1+n_2+4}\frac{\p^2 z_4}{\p p_{n_1}\p p_{n_2}} = 0
\nn
\ee
Already at this level the difference between the two correct formulas looks quite pronounced --
and it only increases at the next levels.
Some new ideas are needed to express the (vanishing) difference in terms of the W-constraints
(\ref{wcons}) and, hopefully, find a concise and universal expression for this discrepancy.
Since it relates two clearly distinguished quantities -- the {\bf SE} which is a single
polynomial equation, which defines $Z\{p_k\}$, and {\bf ME} which is the distinguished
scalar projection of the fundamental matrix Ward-identity (\ref{GN}) --
there {\it should} be some simple relation between them.
We see that the hope of \cite{MMMish}, that this relation is just an identity, fails.
But in the simplest cases (like the basic Kontsevich model $r=2$) it is true --
and thus the discrepancy is an {\it anomaly}, in the sense which still remains to be formulated.
Anyhow, so far anomalies were always comprehensible -- hopefully this will be the case with this
new one as well.

\section{Conclusion}

In this paper we studied the properties of the {\it main equation} (\ref{maineq}) from \cite{MMMish}.
This is important because this equation seems to somehow accumulate the power of
all the $W$-constraints in monomial GKM
and fully define the time dependence of its partition function $Z_r\{p\}$.
In particular it should imply that this partition function
is independent of $p_{rk}$ (of time-variables with the numbers divisible by $r$).
We demonstrated that it does so in an elegant way:
if there was a $p_{rk}$-dependence in $Z_r$, we would not get an equation for it,
which is {\it polynomial} in time-variables $p_k$.
Since $Z_r$ is known from \cite{MMM,GKM} to be a KP $\tau$-function
(this is relatively simple to demonstrate),
independence of $p_{rk}$ means that it belongs to the $r$-reduction of KP hierarchy.
In fact, one can consider our calculation as a new kind of a proof of this statement
(that $Z_r$ is an $r$-{\it reduced} $\tau$-function),
but still a rather sophisticated and undirect one.
A concise, clear and direct proof remains highly desirable.

Also desirable is a direct relation of our calculation with the elegant
description \cite{FKN} of the $W$-constraints for $r$-reductions
as a normal ordering of "circular formula" $\prod_{m=1}^r J\left(z\cdot e^{2\pi i m/r}\right)$.
There are now few doubts that the $W$-constraints are implied by the single main equation --
but the way it works remains unclear.
One can only hope that if this is clarified, the constraints will also come in some clever form --
probably, provided by the circular formula.

Our main result is that the {\it main} equation (\ref{maineq}),
directly following from the matrix Ward identity (\ref{GN}),
is not exactly the same as the "{\it single} equation" (SE) of \cite{MMMR2,MMMish},
but differs from it by additional non-polynomial (!) terms,
which presumably are proportional to

(a) some lower $W$-constraints and

(b) the terms with derivatives over $p_{kr}$, which do not contribute
for $r$-reduced partition functions:

\noindent
for a matrix $M=diag(\mu_1,\ldots,\mu_N)$ of the size $N$ and on the locus $p_k = \sum_{i=1}^N \mu_i^{-k}$
\be
\boxed{
\overbrace{\tr \left( M\left\{\left(\frac{\p}{\p L^{tr}}\right)^r - L\right\}\right)}^{ME} \ = \
\overbrace{-\hat l_0 +  \sum_{i=1}^{r-1} \sum_{n=1}^\infty p_{rn-i} \hat W^{r+1-i}_{n-r-1+i }}^{SE}
\ \
+ \ \ \frac{ O\left(\hat W^{(2)},\ldots, \hat W^{(r-1)}, \frac{\p}{\p p_{kr}}\right)}
{S_{[(r-1),2(r-1), \ldots, (N-1)(r-1)]}}
}
\ee
The fact that
some other $W$ constraints emerge in addition to the {\it single} equation
in the truly-first-principle approach (based on \cite{GN})
can be important for better understanding of its surprising predictive power --
a possibility for a single equation to substitute the entire set of the $W$-constraints
(which has more than one generator: already two, $L_{-1}$ and $L_2$, for $r=2$).
As a byproduct of our calculation we  found an amusing structure of non-polynomial terms,
with a peculiar embedded dependence on $N$.
Since non-polynomial terms are coefficients of $\p Z_r/\p p_{kr}$
which actually vanish for the GKM partition function,
the true significance of these formulas, at least in the case (b), remain unclear --
still they look interesting by themselves and can show up in some other contexts.

The observation of "anomaly" ${\bf SE}\neq {\bf ME}$ even for $r$-reduced $\tau$-functions
leaves the puzzle of $W$-constraints and the origin of $W$-representation for GKM with $r>2$
\cite{MMMish} unsolved.
This adds to the equally puzzling complication of superintegrability formulas
and character calculus for $r>2$: at least the appropriate basis of $Q$-functions
\cite{MMhl} remains unknown.
It is unclear if there is a direct connection between these two complications --
anyhow, the story of GKM is still incomplete and at least one additional idea is still lacking.
Of course the previous ideas, like "circular formula" \cite{FKN}
and non-abelian W-representation \cite{MMMR2,MMMish},
also need to be polished and brought to the same level of clarity
as determinantal representation and KP integrability \cite{GKM}, --
but this is hard to do before the "anomaly" issue is fixed,
which controls the puzzle of $r$-reduction  and the very origin of
sophisticated $W$-constraints and the way they follow from
the apparent original Ward identity \cite{GN}.
If superintegrability and character expansion will also get clarified by the resolution
of this puzzle, or need to wait for additional insights, remains to be seen.

Last but not the least -- all the formulas in this paper
are obtained for particular low values of $r$ and $N$,
what is enough to reveal the emerging structures and phenomena.
Still general consideration and proofs remain to be given.
They can also bring new ideas and further develop and clarify the theory of GKM,
which remains mysteriously complicated and transcendent -- perhaps a little less now,
but still far from simplicity and transparency achieved for the other eigenvalue matrix models
(including the cubic GKM).

\section*{Acknowledgements}

Nearly 45 years of cooperation with Andrei were precious.
This paper concerns just a small unclear piece in one of the branches of our common interest.

I am indebted to A.Mironov, V.Mishnyakov and A.Zhabin
for fresh discussions of GKM and related subjects.

This work is supported by the Russian Science Foundation (Grant No.21-12-00400).


\begin{thebibliography}{12}


\bibitem{Kon} M.~Kontsevich,
  Commun.Math.Phys.\  {\bf 147} (1992) 1

\bibitem{MMM} A.~Marshakov, A.~Mironov, A.~Morozov,
  Phys.Lett. {\bf B274} (1992) 280,

\bibitem{W} E.Witten, {\sl On the Kontsevich model and other models of
two-dimensional gravity}, in: New York 1991 Proc., Differential geometric
methods in theoretical physics, v.1, pp.176-216



\bibitem{GKM} S.~Kharchev, A.~Marshakov, A.~Mironov, A.~Morozov, A.~Zabrodin,
  Phys.Lett. {\bf B275} (1992) 311,
  hep-th/9111037\\
S.~Kharchev, A.~Marshakov, A.~Mironov, A.~Morozov, A.~Zabrodin,
  Nucl.Phys.\ {\bf B380} (1992) 181,
  hep-th/9201013
  
\bibitem{versus} S. Kharchev, A. Marshakov, A. Mironov, A. Morozov,
Nucl.Phys. {\bf B397} (1993) 339-378, hep-th/9203043



  
\bibitem{UFN3} A. Morozov,
Phys.Usp.(UFN) {\bf 37} (1994) 1, hep-th/9303139;
hep-th/9502091; hep-th/0502010\\
A. Mironov, Int.J.Mod.Phys. {\bf A9} (1994) 4355; Phys.Part.Nucl.
{\bf 33} (2002) 537; hep-th/9409190

\bibitem{AMMP} A.~Alexandrov, A.~Mironov, A.~Morozov, P.~Putrov,
  Int.J.Mod.Phys. {\bf A24} (2009) 4939,
arXiv:0811.2825  

\bibitem{Zhou} J.Zhou,
arXiv:1305.6991
  
\bibitem{MMhl} A.Mironov and A.Morozov,  
EPJ C 81 (2021) 270, arXiv:2011.12917;
Phys.Lett. B 819 (2021) 136474,  arXiv:2101.08759 \\
A. Alexandrov, arXiv:2012.07573  \\
A.Mironov, A.Morozov, A.Orlov, S.Natanzon,  arXiv:2012.09847 \\
X.Liu, Ch.Yang, 
arXiv:2103.14318,
arXiv:2104.01357;
{\it Action of Virasoro operators on Hall-Littlewood polynomials}, to appear

     
 
  
 

\bibitem{MMMR1} A.~Mironov, V.~Mishnyakov, A.~Morozov, R.~Rashkov,
JETP Letters {\bf 113:11} (2021),
arXiv:2104.11550

\bibitem{MMMR2} A.~Mironov, V.~Mishnyakov, A.~Morozov, R.~Rashkov,
arXiv:2105.09920

\bibitem{MMMish} A.~Mironov, V.~Mishnyakov, A.~Morozov, 
arXiv:2107.02210    



\bibitem{GN} D. Gross, M. Newman,
Nucl.Phys. {\bf B380} (1992) 168-180 





\bibitem{FKN} M.~Fukuma, H.~Kawai, R.~Nakayama,
  Int.\ J.\ Mod.\ Phys.\ {\bf A6} (1991) 1385;
Comm.Math.Phys. {\bf 143} (1992) 371-403


\bibitem{wrep0} A. Givental, 
math.AG/0008067\\
A.Okounkov,
Math.Res.Lett. {\bf 7}
(2000) 447-453;\\
A. Alexandrov, A. Mironov, A. Morozov,
Physica {\bf D235} (2007) 126-167, hep-th/0608228;
Theor. Math. Phys. \textbf{150} (2007) 153-164,
hep-th/0605171 \\
V.Bouchard, M.Marino,
In: {\sl From Hodge Theory to Integrability and tQFT: tt*-geometry},
Proceedings of Symposia in Pure Mathematics, AMS (2008), arXiv:0709.1458;\\
S.Lando,
In: {\sl Applications of Group Theory to Combinatorics}, Koolen,
Kwak and Xu, Eds.
Taylor \& Francis Group, London, 2008, 109-132;\\
M.Kazarian,
arXiv:0809.3263;\\
A.Mironov, A.Morozov,
JHEP \textbf{0902} (2009) 024, arXiv:0807.2843\\
A.Mironov, A.Morozov and G.Semenoff,  Int.J.Mod.Phys. A11 (1996) 5031, hep-th/9404005 \\
A.Mironov, A.Morozov and S.Natanzon, 
Theoretical and Mathematical Physics,  166, no. 1 (2011)  1–22 \\ 
A. Mironov, A. Morozov, A. Zhabin, 
 arXiv:2105.10978     

\bibitem{Wrep} A.~Morozov, S.~Shakirov,
  JHEP {\bf 0904} (2009) 064,
arXiv:0902.2627 \\
A. Alexandrov, Mod.Phys.Lett. {\bf A26} (2011) 2193-2199, arXiv:1009.4887; 
Adv.Theor.Math.Phys. 22 (2018) 1347, arXiv:1608.01627 \\
H. Itoyama, A. Mironov, A. Morozov, JHEP 1706 (2017) 115, arXiv:1704.08648 \\
A. Mironov, A. Morozov, Phys. Lett. B 771 (2017) 503, arXiv:1705.00976  




\bibitem{vircon}
A. Mironov, A. Morozov, Phys.Lett. {\bf B252} (1990) 47-52\\
F. David, Mod.Phys.Lett. {\bf A5} (1990) 1019\\
J. Ambj{\o}rn, Yu. Makeenko, Mod.Phys.Lett. {\bf A5} (1990) 1753\\
H. Itoyama, Y. Matsuo, Phys.Lett. {\bf 255B} (1991) 20 \\
A.~Alexandrov, A.~Mironov, A.~Morozov,
  Int.\ J.\ Mod.\ Phys.\ A {\bf 19} (2004) 4127,
hep-th/0310113 \\
R.Lodin, A.Popolitov, Sh.Shakirov and M.Zabzine
Lett Math Phys 110 (2020) 179-210,  
 arXiv:1810.00761 



\bibitem{LL3} L.D.Landau, E.M.Lifshitz, {\it  Quantum Mechanics: Non-Relativistic Theory. Vol. 3},
  (1958) Pergamon Press

\bibitem{Mikh} A.~Mikhailov,
  Int.\ J.\ Mod.\ Phys.\ {\bf A9} (1994) 873,
  hep-th/9303129
  
  
\bibitem{CheMa} L.Chekhov and Yu.Makeenko,
 Phys.Lett. 278B (1992) 271, hep-th/9202006  





 

 

 


 


 
\end{thebibliography}
\end{document}